\documentstyle[12pt,epsfig]{article}

\def\bc{\begin{center}}
\def\ec{\end{center}}
\def\beq{\begin{equation}}
\def\eeq{\end{equation}}

\def\at#1{\left. \right|^{}_{#1}}
\def\hs#1{\hspace*{#1cm}}

\def\av#1{\langle {#1} \rangle}
\def\avr#1#2{\langle {#1} \rangle^{}_{#2}}
\def\etai{\sqrt{\eta_i}}
\def\F{{n^{}_F}}
\def\B{{n^{}_B}}

\def\ppB{{p^{2}_{{\rm t}B}}}
\def\Ceta{{\{\eta^{}_1,...,\eta^{}_M\}}}
\def\Aeta{{\eta^{}_1,...,\eta^{}_M}}
\def\Cn{{\{n^{}_1,...,n^{}_M\}}}

\def\bfs{{\bf s}}
\def\olb{\overline{b}}
\def\olbe{\overline{\beta}}
\def\olx{\overline{x}}

\def\olN{\overline{N}}
\def\olG{\overline{G}}
\def\olP{\overline{P}}
\def\oln{\overline{n}}
\def\bN{\overline{N}}
\def\br{\overline{r}}
\def\bn{\overline{n}}

\def\ol#1{\overline{#1}}
\def\beti{\ol{\eta}^{}_{i}}
\def\reta{\sum_{i}^{} \sqrt{\eta_i}}

\def\seta{\sum_{i}^{} \eta_i}

\def\j#1#2#3#4{{#1} {\bf #2}, #4 (#3)}

\def\NP{Nucl. Phys.}
\def\PL{Phys. Lett.}
\def\PRL{Phys. Rev. Lett.}

\def\ZP{Z. Phys.}
\def\PRep{Phys. Rep.}

\def\EPJ{Eur. Phys. J.}
\def\IJMP{Int. J. of Mod. Phys.}

\title{
Cellular Approach to Long-Range\\
$p_t$ and Multiplicity Correlations\\
in the String Fusion Model
}
\author{
{\bf V.V. Vechernin and R.S. Kolevatov}\\
{\it High Energy Physics Dep., St. Petersburg State University,}\\
{\it 1 Ulianovskaya str., 198504 St.Petersburg, Russia}
}
\date{}

\begin{document}

\begin{titlepage}
\maketitle
\medskip


\begin{abstract}

The long-range $p_t$ and multiplicity($n$) correlations
in high-energy nuclear collisions
are studied in the framework of a
simple cellular analog of the string fusion model.

Two cases with local and global string
fusion is considered.
The $p_t$--$n$ and $n$--$n$ correlation functions and
correlation coefficients are calculated analytically
in some asymptotic cases
using suggested Gauss approximation.

It's shown that at large string density
the $p_t$--$n$ and $n$--$n$ correlation coefficients
are connected and the scaling takes place.
The behavior of the correlations at small string density
is also studied.

The asymptotic results are compared with results
of the numerical calculations in the framework of
proposed cellular approach.

\end{abstract}
\vspace*{1cm}
\end{titlepage}


\section{Introduction.}

The colour strings approach \cite{Capella1,Kaidalov}
is widely applied for the description of the soft part of
the hadronic and nuclear interactions at high energies.

In the frame work of this approach
the string fusion model was suggested in papers \cite{BP1}.
Later it was developed \cite{ABP1}-\cite{BPR} and applied for the
description of the long-range multiplicity and $p_t$ correlations
in relativistic nuclear collisions \cite{PRL94}-\cite{BP00}.

In the paper \cite{vestn1} we have
formulated some simple cellular analog of the
string fusion model,
which enables explicit analytical calculations
of the correlation functions in some asymptotic cases and
can simplify calculations in the case of real nuclear
collisions.

In that paper we have checked up the assumptions
of the cellular approach
and the validity of a suggested Gauss approximation
in the simplest (no fusion)
case when the explicit solution of the model can be found.

In the present paper
in the framework of proposed cellular approach
we calculate
$p_t$--$n$ and $n$--$n$ correlation functions and
correlation coefficients
in two cases: with local and with global string
fusion.

The calculations are done both
numerically
and in some asymptotic cases
analytically
using the Gauss approximation.
The results of both calculations are in a good agreement,
which proofs the validity of proposed Gauss approximation.

At large string density $\eta \gg 1$
the connection between
the $p_t$--$n$ and $n$--$n$ correlation coefficients
are found in both cases: with local or global string
fusion.
At that
for the correlation coefficient
the scaling takes place.
It depends only on one combination $\mu_0/\sqrt{\eta}$
of the variables $\eta$ and $\mu_0$ ($\mu_0$ is
the mean multiplicity emitting by a single string).

The paper organized as follows.
In the next section we recall
the formulation of the cellular approach
in the case with a {\it local} string fusion.

In the section 3 we develop Gauss approximation
at large string density $\eta$ in the {\it local} fusion case.

In the section 4
the $p_t$--$n$ and $n$--$n$ correlation coefficients
are calculated at large $\eta$ in the case with {\it local} string fusion
and connection between these two coefficients is found.

In the section 5
the $p_t$--$n$ and $n$--$n$ correlation functions and
correlation coefficients
are calculated at large $\eta$ for "homogeneous" situation
(a constant mean density of strings)
in the case with {\it local} fusion.
It's shown that in this situation
the $p_t$--$n$ and $n$--$n$ correlation coefficients become equal
and the $\mu_0/\sqrt{\eta}$-scaling takes place.
The obtained results are compared with the results
of numerical calculations using formulas of section~2.

In the section 6 we recall
the formulation of the cellular approach
in the case with a {\it global} string fusion.
The exact closed formulas for
the $p_t$--$n$ and $n$--$n$ correlation functions
in this case are obtained.

In the section 7 we develop Gauss approximation
at large string density $\eta$ for the {\it global} fusion case.
The $p_t$--$n$ and $n$--$n$ correlation functions and coefficients
for this case are calculated and compared with the results
of numerical calculations using formulas of section~6.

The behavior of the correlations at small string density
is studied in Appendix A.


\section{Cellular approach to the local string fusion.}

Let us recall the formulas obtained for this case in \cite{vestn1}.
We consider the collision of nuclei in two stage scenario
when at first stage the colour strings are formed,
and at the second stage these strings,
or some other (higher colour) strings
formed due to fusion of primary strings, are decaying, emitting
observed particles.

In principle, one can consider two types of fusion.
The case with a local fusion corresponds to the model,
where colour fields
are summing up only locally and the global fusion case corresponds
to the model, where colour fields are summing up globally -
all over the cluster area - into one average colour field,
the last case corresponds to the summing of the sources colour charges.
(In section 5 of \cite{vestn1}
we have referred to these cases as A) and B) correspondingly.)

In the transverse plane depending on the impact
parameter $b$ we have some interaction area $S(b)$.
Let us split this area on the cells
of order of the transverse string size.
Then we'll have $M=S(b)/\sigma_0$ cells,
where $\sigma_0=\pi r_0^2$ is the transverse area of the string
and $r_0\approx 0.2 fm$ is the string radius.

In the case with a local fusion the assumption of the model is
that if the number of strings
belonging to the $i$-th cell is $\eta_i$, then they form
higher colour string, which emits in average $\mu _0\sqrt{\eta _i}$ particles
with mean $p_t^2$ equal to $p^2\sqrt{\eta _i}$,
compared with $\mu _0$ particles with $\langle p_t^2 \rangle =p^2$
emitting by a single string.

If we denote by $n_i$ and $\bn_i$ - the number and the average
number of particles emitted by the higher string from $i$-th cell
in a given rapidity interval, then
\beq
\bn_i=\mu _0\sqrt{\eta _i}
\label{fus1}
\eeq

From event to event the number of strings $\eta_i$ in $i$-th cell
will fluctuate around some
average value - $\overline{\eta }_i^{}$.
Clear that in the case of real nuclear collisions
these average values $\overline{\eta }_i^{}$ will be different
for different cells. They will depend on the position (${\bf s}$)
of the $i$-th cell in the interaction area
(${\bf s}$ is two dimensional vector in transverse plane).
To get the physical answer we have to sum the contributions
from different cells, which corresponds to integration over
${\bf s}$ in transverse plane.

The average local density of primary strings $\overline{\eta }_i^{}$
in the point ${\bf s}$ of transverse plane is uniquely
determined by the distributions of nuclear densities and
the value of the impact parameter - $b$.
They can be calculated, for example, in Glauber approximation.
We'll do this later in a separate paper. In present paper
we consider that all $\overline{\eta }_i^{}$ are already
fixed from these considerations at given value
of the impact parameter - $b$.

Let us introduce some quantities, which will play important
role in our consideration:
$$
N=\sum_{i=1}^M \eta_i , \hs{1} \bN=\sum_{i=1}^M \beti
$$
\beq
r=\sum_{i=1}^M \sqrt{\eta_i} , \hs{1} \br=\sum_{i=1}^M \sqrt{\beti}
\label{not}
\eeq
then clear that $N$ is the number of strings in the given event
and $\bN$ is the mean number of strings for this type of events
(at the fixed impact parameter $b$).

To go to long-range rapidity correlations we have to consider
two rapidity windows $F$ (forward) and $B$ (backward).
Each event corresponds to a certain configuration $\Ceta$ of
strings and certain numbers of charged particles $\Cn$ emitted by
these strings in the forward rapidity window. Then the total
number of particles produced in the forward rapidity window
will be equal to $n_F$:
\beq
n_F=\sum_{i=1}^{M} n_i
\label{nF}
\eeq
The probability to detect
$n_F$ particles in the forward rapidity window
for a given configuration $\Ceta$ of strings
is equal to
\beq
P^{}_{\{\eta^{}_1,...,\eta^{}_M\}} (\F)=
\sum_{\{n^{}_1,...,n^{}_M\}} \delta_{\F,\sum_i n_i} \prod_{i=1}^M
p^{}_{\eta_i} (n_i)
\label{PnF}
\eeq
where $p^{}_{\eta_i} (n_i)$ is the probability of the emission
of $n_i$ particles by the string $\eta_i$ in the forward
rapidity window. By our assumption (\ref{fus1})
\beq
\bn_i\equiv\sum_{n_i=0}^{\infty} n_i p^{}_{\eta_i} (n_i) =
\mu_0 \sqrt{\eta_i}
\label{fusion}
\eeq

If we denote else by $W(\Aeta)$ the probability of realization of the
string configuration $\Ceta$ in the given event,
then the average value of some quantity $O$ under condition of
the production of $n_F$ particles in the forward window
will be equal to
\beq
\avr{O}{\F}
=\frac{\sum_{\{\eta^{}_1,...,\eta^{}_M\}} \avr{O} {{\{\eta^{}_1,...,\eta^{}_M\}},\F} W(\Aeta) P^{}_{\{\eta^{}_1,...,\eta^{}_M\}} (\F)}
{\sum_{\{\eta^{}_1,...,\eta^{}_M\}} W(\Aeta) P^{}_{\{\eta^{}_1,...,\eta^{}_M\}} (\F)}
\label{avO}
\eeq
One has to omit in this $M$-fold sums one term, when all $\eta_i=0$,
which corresponds to the absence of inelastic interaction between
the nucleons of the colliding nuclei (see details in Appendix A).

If the $O$ in the number of particles produced in the backward
rapidity window $n_B$ in the given event, then
we have to use for $\avr{\B}{\F}$ correlations:
\beq
\avr{\B} {{\{\eta^{}_1,...,\eta^{}_M\}},\F} =
\mu_0 \sum_{i=1}^{M} \sqrt{\eta_i}=\mu_0 r
\label{Onn}
\eeq
If the $O$ in the mean squared transverse momentum
of particles produced in the backward
rapidity window $\ppB$ in the given event, then
we have to use for $\avr{\ppB}{\F}$ correlations:
\beq
\avr{\ppB} {{\{\eta^{}_1,...,\eta^{}_M\}},\F}
= \sum_{i=1}^{M} \frac{\sqrt{\eta_i}}
{\sum_{i=1}^{M} \sqrt{\eta_i}} p^2 \sqrt{\eta_i}
= p^2 \frac{\sum_{i=1}^{M} \eta_i}{\sum_{i=1}^{M} \sqrt{\eta_i}}
=p^2\frac{N}{r}
\label{Optn}
\eeq

Later we'll assume that numbers of primary strings in each cell $\eta_i$
fluctuate independently around some average quantities
$\overline{\eta }_i^{}$ uniquely
determined by the distributions of nuclear densities and
the value of the impact parameter - $b$ (see above), then
\beq
W(\Aeta) = \prod_{i=1}^M w (\eta_i), \hs{1}
\sum_{i=1}^M \eta_i w (\eta_i) = \beti
\label{WCeta}
\eeq

For clearness we'll sometimes address to a simple
"homogeneous" case, when all $\beti$
(but not the $\eta_i$, which fluctuate from event to event!)
is equal each other in the interaction area $\beti=\eta$
(a constant mean density of strings).
The parameter $\eta$ coincides in this case with
the parameter $\eta$ used in the papers \cite{BPR,BPep00,BP00}
and has the meaning of the mean number of strings
per area of one string ($\eta=({\rm mean\ string\ density})\times\sigma_0$).
In general case the parameters $\beti$ have the same meaning,
but with mean string density depending on the point $\bfs$
in the transverse interaction plane\\
($\beti=({\rm mean\ string\ density\ in\ the\ point}\ \bfs)\times\sigma_0$).

As it was shown in \cite{vestn1}
if we assume else the Poissonian form of $p^{}_{\eta_i} (n_i)$
($\rho^{}_{a}(x)$ is the Poisson distribution with $\olx=a$):
\beq
p^{}_{\eta_i} (n_i) =  \rho^{}_{\mu_0\etai}(n_i)
 \equiv e^{-\mu_0\etai}\frac{(\mu_0\etai)^{n_i}}{n_i!}
\label{Poisson}
\eeq
then we find the Poissonian distribution for
\beq
P^{}_{\{\eta^{}_1,...,\eta^{}_M\}} (\F) = \rho^{}_{\mu_0\reta}(\F)
\label{Poisso}
\eeq
with $\avr{\F} {{\{\eta^{}_1,...,\eta^{}_M\}}} = \mu_0 \reta = \mu_0 r = \avr{\F} {r}$
and $\sigma_{rF}^2=\avr{\F} {r}=\mu_0 r$, which we can
replace at large $\mu_0 \reta$ by Gauss distribution:
\beq
P^{}_{\{\eta^{}_1,...,\eta^{}_M\}} (\F)
= \frac{1}{\sqrt{2\pi}\sigma_{rF}}
e^{-\frac{(\F-\avr{\F} {r})^2}{2\sigma_{rF}^2}}
\label{pgausA}
\eeq

It was also shown in the section 6 of \cite{vestn1} that
if we assume the Binomial form of $p^{}_{\eta_i} (n_i)$
then we find the Binomial distribution for $P^{}_{\{\eta^{}_1,...,\eta^{}_M\}} (\F)$
with $\avr{\F} {{\{\eta^{}_1,...,\eta^{}_M\}}} = \mu_0 \reta = \mu_0 r = \avr{\F} {r}$
and $\sigma_{rF}^2=\avr{\F} {r} (1-\lambda)=\mu_0 r (1-\lambda)$,
where $\lambda\to 0$ corresponds to the Poisson limit and
$\lambda\to 1$ corresponds to the case when each higher string emits
fixed number of particles - $n_i=\oln_i=\mu_0 \sqrt{\eta_i}$
in each event.

Moreover appealing to the central limit theorem of the probability theory
one can state that at large $M$
we'll have formula (\ref{pgausA}) for any type of $p^{}_{\eta_i} (n_i)$.

\section {Gauss approximation
 at large string density for the {\it local} fusion.}

Now we'll evaluate $\sum_{\{\eta^{}_1,...,\eta^{}_M\}}$ at the condition that all $\beti\gg 1$.
At this condition we can use Gauss approximation for each $w (\eta_i)$
\beq
w (\eta_i)
= \frac{1}{\sqrt{2\pi}\sigma_{\eta_i}}
e^{-\frac{(\eta_i-\beti)^2}{2\sigma_{\eta_i}^2}}
\label{wgausA}
\eeq
with  $\sigma_{\eta_i}^2=\beti (1-\lambda_\eta)$,
where again $\lambda_\eta\to 0$ corresponds to the Poisson limit and
$\lambda_\eta\to 1$ corresponds to the case with a fixed number of strings
$N=\bN=\sum_{i=1}^M \beti$ (see the section 6 of \cite{vestn1}).

As in \cite{vestn1} we have then
\beq
\avr{\B} {\F}
=\mu_0 \frac{\int d\eta^{}_1...d\eta^{}_M r \frac{1}{\sqrt{r}}e^{-\varphi(\eta_i,\F)}}
{\int d\eta^{}_1...d\eta^{}_M \frac{1}{\sqrt{r}}e^{-\varphi(\eta_i,\F)}}
\label{BF1A}
\eeq
and
\beq
\avr{\ppB} {\F}
=p^2 \frac{\int d\eta^{}_1...d\eta^{}_M \frac{N}{r} \frac{1}{\sqrt{r}}e^{-\varphi(\eta_i,\F)}}
{\int d\eta^{}_1...d\eta^{}_M \frac{1}{\sqrt{r}}e^{-\varphi(\eta_i,\F)}}
\label{pBF1A}
\eeq
Recall that $N=\sum_{i=1}^M \eta_i$ and $r=\sum_{i=1}^M \sqrt{\eta_i}$. Here
\beq
\varphi(\eta_i,\F)=\sum_{i=1}^M\frac{(\eta_i-\beti)^2}{2\beti(1-\lambda_\eta)}
+\frac{(\F-\mu_0 r)^2}{2\mu_0 r(1-\lambda)}
\label{phiA}
\eeq

Further we take out factors before exponent
at the point, where $\varphi$ is minimal, after that
the rest integrals in the numerator and in the denominator are being reduced, and we find
\beq
\avr{\B} {\F} =\mu_0 r^*
\label{BF2A}
\eeq
and
\beq
\avr{\ppB} {\F} =p^2 \frac{N^*}{r^*}
\label{pBF2A}
\eeq
The $N^*$ and $r^*$ are the values of $N$ and $r$ in the point
$\{\eta^{*}_1,...,\eta^{*}_M\}$, where $\varphi(\eta_i,\F)$ is minimal:
\beq
\frac{\partial\varphi(\eta_i,\F)}{\partial\eta_i}=0
\label{dphiA}
\eeq
This leads to the system of equations:
\beq
\frac{\eta_i^*}{\beti}-1= \frac{\mu_0\kappa}{4\sqrt{\eta_i^*}}
\left( \frac{n_F^2}{\mu_0^2 r^{*2}}-1 \right)
\label{etistarA}
\eeq
Recall that $r^*=\sum_{i=1}^M \sqrt{\eta_i^*}$ and
\beq
\kappa=\frac{1-\lambda_\eta}{1-\lambda}
\label{kappa}
\eeq
For the meaning of $\kappa$ see
the end of the previous section and
the section 6 of \cite{vestn1},
for both Poissonian distributions $\kappa=1$
and $\kappa$ is the relative width of the
$p(n_i)$ and $w(\eta_i)$ distributions in other cases.
The (\ref{etistarA}) defines $\eta_i^*$ as function of $\F$.

Introducing short notations
\beq
z_i=\sqrt{\frac{\eta_i^*}{\beti}}, \hs{1}
f=\frac{\F}{\mu_0 \br}=\frac{\F}{\av{\F}}, \hs{1}
a_i=\frac{\mu_0\kappa}{4\sqrt{\beti}}
\label{notA}
\eeq
we can rewrite (\ref{etistarA}) as
\beq
z_i^3-z_i^{}=a_i \left( f^2 \frac{\br^2}{r^{*2}}-1 \right)
\label{ziA}
\eeq
where $\br=\sum_{i=1}^M \sqrt{\beti}$, $r^*=\sum_{i=1}^M z_i\sqrt{\beti}$
and $N^*=\sum_{i=1}^M z_i^2 \beti$.
The (\ref{ziA}) defines $z_i$ as function of $f$: \
$z_i=z_i(f)$. Then we can calculate
$\avr{\B} {\F}$ and $\avr{\ppB} {\F}$ as a function of $\F$
using (\ref{BF2A}) and (\ref{pBF2A}).


\section {$p_t$--$n$ and $n$--$n$ correlation coefficients 
 at a large  string density for the {\it local} fusion.}

The correlation coefficients are defined in the same way as in
the section 4 of \cite{vestn1}:
\beq
b \equiv \frac{d\avr{\B} {\F}}{d\F} \at {\F=\av{\F}}
\label{bA}
\eeq
and
\beq
\beta \equiv \frac{d\avr{\ppB} {\F}}{d\F} \at {\F=\av{\F}}
\label{betaA}
\eeq
or for "relative" quantities
\beq
\ol{b} \equiv\frac{d\avr{\B} {\F}/\av{\B}}{d\F/\av{\F}} \at {\F=\av{\F}}
\label{bbA}
\eeq
and
\beq
\ol{\beta} \equiv\frac{d\avr{\ppB}{\F}/\av{\ppB}}{d\F/\av{\F}} \at {\F=\av{\F}}
\label{bbetaA}
\eeq
(Note the same definition of
$pt$--$n$ correlation coefficient $\ol{\beta}$
in \cite{Heiselberg} (see formula (44) in \cite{Heiselberg}),
see also remark in Appendix B).

In short notation using (\ref{BF2A}) and (\ref{pBF2A}) we have:
\beq
\ol{b} =\frac{1}{\br}\frac{dr^*}{df} \at {f=1}
\label{bb1A}
\eeq
and
\beq
\ol{\beta} =\frac{\br}{\bN}\frac{d(N^*/r^*)}{df} \at {f=1}
\label{bbeta1A}
\eeq

We can't solve the equations (\ref{ziA}) for to find $z_i=z_i(f)$ explicitly,
but to calculate the correlation coefficients we need to know
only $z'_i(1)=\frac{dz_i(f)}{df}\at {f=1}$, which can be done explicitly.

We see that at $f=1$ (\ref{ziA}) has the obvious solution:
\beq
f=1, \hs{1} z_i=1, \hs{1} \eta^*_i=\beti, \hs{1} r^*=\br, \hs{1} N^*=\bN
\label{solziA}
\eeq
We need to calculate $z'_i(f)$ only at $f=1$. Differentiating
(\ref{ziA}) on $f$ and using then again (\ref{solziA}) we find
\beq
z'_i(1)=a_i \frac{4\br}{4\br+\mu_0\kappa M}
\label{solzi1A}
\eeq
with $a_i=\mu_0\kappa/(4\sqrt{\beti})$. Then
\beq
\ol{b} =\frac{1}{\br}\frac{dr^*}{df} \at {f=1}=\frac{1}{\br}\sum_{i=1}^M z'_i(1)\sqrt{\beti}
=\frac{\mu_0\kappa}{\mu_0\kappa+4\br/M}
\label{bb2A}
\eeq
and
\beq
\ol{\beta} =\frac{1}{\bN} \frac{dN^*}{df} - \frac{1}{\br} \frac{dr^*}{df} \at {f=1}
=\frac{1}{\bN} \frac{dN^*}{df} \at {f=1} -\ol{b}
\label{bbeta2A}
\eeq
Using
\beq
\frac{dN^*}{df} \at {f=1}=2\sum_{i=1}^M z'_i(1)\beti
=\frac{2\mu_0\kappa\br^2}{\mu_0\kappa M+4\br}
\label{dNdfA}
\eeq
we have
\beq
\ol{\beta}
=  \left( \frac{2\br^2}{\bN M}-1  \right)
\frac{\mu_0\kappa}{\mu_0\kappa+4\br/M}
=  \left( \frac{2\br^2}{\bN M}-1  \right) \ol{b}
\label{bbeta3A}
\eeq
We see the connection between $pt$--$n$ and $n$--$n$ correlation coefficients.
Note that due to obvious inequality:
\beq
\left( \sum_{i=1}^M \sqrt{\beti} \right)^2  \leq  M \sum_{i=1}^M \beti
\label{ineqv}
\eeq
we have $\br^2 \leq M \bN$ and hence always $\ol{\beta}\leq\ol{b}$.

Clear that at equal $\beti\equiv\eta$ we have
$\br=M\sqrt{\eta}$, $\bN=M\eta$, $\br^2 = \bN M$ and
\beq
\ol{\beta}=\ol{b}
=\frac{\mu_0\kappa}{\mu_0\kappa+4\sqrt{\eta}}
\label{bbb3A}
\eeq

From (\ref{bb2A}) we see that $n$--$n$ correlation coefficient
is always positive.
Can $pt$--$n$ correlation coefficient be negative?
Let us consider {\it nonhomogeneous} situation when
$\beti=\eta_+$ at $i=1,...,M_1$,
$\beti=\eta_-$ at $i=M_1+1,...,M$, \  $M_1\sim M$ and $\eta_+ \gg \eta_- \gg 1$.
Then
$\br=M_1\sqrt{\eta_+}+(M-M_1)\sqrt{\eta_-}\approx M_1\sqrt{\eta_+}$, \
$\bN=M_1\eta_+ +(M-M_1)\eta_- \approx M_1 \eta_+$ and we have
\beq
\ol{\beta}
\approx  \left( 2\frac{M_1}{M}-1  \right) \ol{b}
\label{bbeta4A}
\eeq
We see that at $M_1<M/2$ we can have $\ol{\beta} < 0 $.


\section{The $\mu_0/\eta^{1/2}$-scaling at large string density.}

Let us consider for clearness
{\it homogeneous} case, when all $\beti$
is equal each other in the interaction area $\beti=\eta$
(a constant mean density of strings).
In this case we can explicitly calculate
at large string density $\eta$
not only the $p_t$--$n$ and $n$--$n$ correlation coefficients,
but also the corresponding correlation functions
for the version with a local string fusion.

We have seen in the end of previous section that in this case
the coefficients for $n$--$n$ and $pt$--$n$ correlations
defined as (\ref{bbA}) and (\ref{bbetaA})
are coincide:
\beq
\ol{\beta}=\ol{b}
=\frac{\mu_0\kappa}{\mu_0\kappa+4\sqrt{\eta}}=\frac{a}{a+1}
\label{bb3A}
\eeq
where $a=\mu_0\kappa/(4\sqrt{\eta})$.

In this homogenous case $\beti\equiv\eta$ we can also calculate
the correlation functions $\avr{\ppB}{\F}$ and $\avr{\B}{\F}$ at any $\F$.
Due to symmetry, the system of equations (\ref{ziA})
have symmetrical solution $z_i=z$ and can be reduced to one equation
\beq
z^3-z=a \left( \frac{f^2}{z^2}-1 \right)
\label{zA}
\eeq
because $\br= M \sqrt{\eta}$,  $r^*=z M \sqrt{\eta}$ and $N^*=z^2 M \eta$ with
\beq
f=\frac{\F}{\mu_0 M \sqrt{\eta}}=\frac{\F}{\av{\F}}, \hs{1}
a=\frac{\mu_0\kappa}{4\sqrt{\eta}}
\label{not1A}
\eeq
The (\ref{zA}) defines the function $z=z(f)$ and then
using (\ref{BF2A}) and (\ref{pBF2A}) we can calculate
\beq
\avr{\B}{\F} =\mu_0 r^*= \mu_0 M \sqrt{\eta} z(f)=\av{\B} z(f)
\label{BF3A}
\eeq
and
\beq
\avr{\ppB}{\F} =p^2 \frac{N^*}{r^*}=p^2 \sqrt{\eta} z(f)=\av{\ppB} z(f)
\label{pBF3A}
\eeq
So we have at any $\F=\av{\F} f $:
\beq
\frac{\avr{\B}{\F}}{\av{\B}}
=\frac{\avr{\ppB}{\F}}{\av{\ppB}}= z(f)
\label{pnA}
\eeq

From (\ref{bb3A}) and (\ref{zA}) we see that
in this homogenous case
at large string density $\eta$
there is an remarkable {\it scaling}.
The $pt$--$n$ and $n$--$n$ correlation coefficients
and correlation functions depend only on
one combination $a=\mu_0\kappa/(4\sqrt{\eta})$
of parameters.

\begin{figure}[t]
\centerline{\epsfig{file=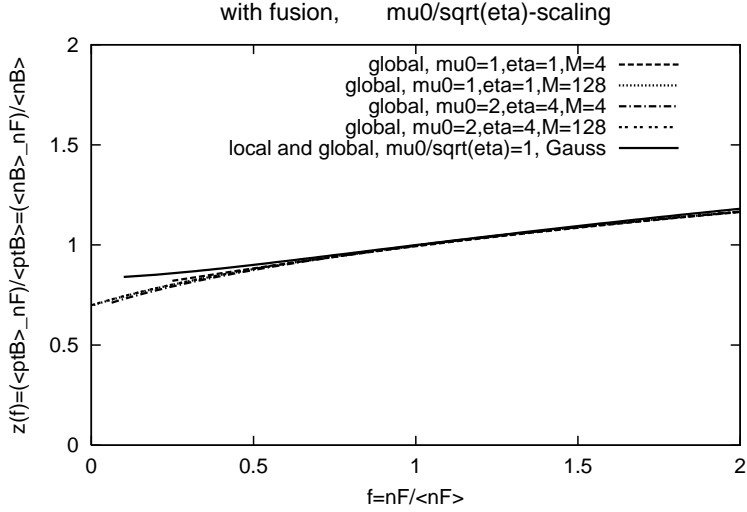,width=10cm,angle=-90}}
\caption[dummy]{\label{zf1}
The $p_t$--$n$ and $n$--$n$ correlation functions.
Solid line - the Gauss approximation for local and global fusion
at the value of scaling variable $\mu_0/\sqrt{\eta}=1$.
Dashed and dotted lines - the results of exact calculations
in the global fusion case at different values of
$\mu_0$, $\eta$ and $M$.
The $\mu_0/\sqrt{\eta}$-scaling and $M$-independence.
}
\end{figure}
\begin{figure}[t]
\centerline{\epsfig{file=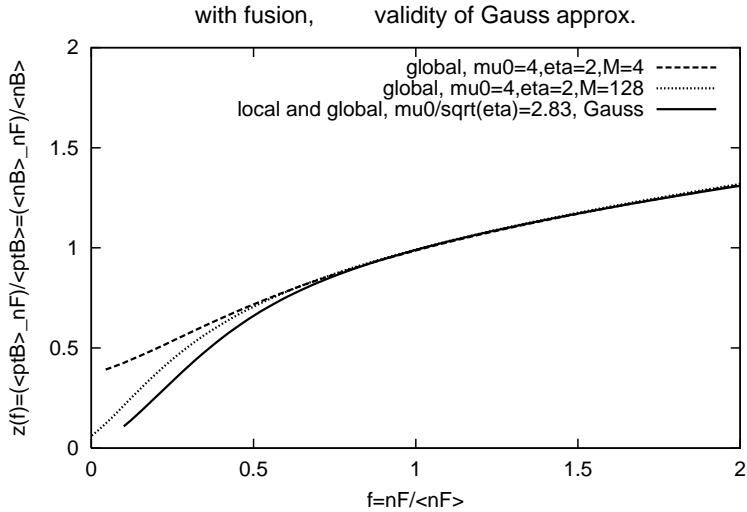,width=10cm,angle=-90}}
\caption[dummy]{\label{zf2}
The same as in Fig.~\ref{zf1}, but at different values of
$\mu_0$ and $\eta$.
}
\end{figure}

We present the correlation function $z(f)$ (\ref{pnA})
in Figs.~\ref{zf1},\ref{zf2}
and the correlation coefficient $\ol{\beta}=\ol{b}$ (\ref{bb3A})
as function of $\eta$ in Figs.~\ref{b1}-\ref{b4}
(the solid lines).
We present also in Figs.~\ref{b1}-\ref{b4}
the results of our direct numerical MC calculations
of the $pt$--$n$ and $n$--$n$ correlation coefficients
in the local fusion case based on formulas (\ref{avO}-\ref{Optn})
(empty and filled points correspondingly).

We see that in the case with local fusion
at small string density we have large
$n$--$n$ correlations (the same as in the case without
string fusion \cite{vestn1}) and no $p_t$--$n$ correlations.
(The analysis at very small values of $\eta\leq 1/M$
see in Appendix A.)

At large string density in the homogeneous case
the $p_t$--$n$ and $n$--$n$ correlation coefficients become equal
and the $\mu_0/\sqrt{\eta}$-scaling takes place.
We see also that in this limit our
Gauss asymptotic is in a good agreement with
results of the numerical calculations and $M$-independence
takes place.


\section{The {\it global} fusion at large string density.
 Exact solution.}

In this case at first stage
we also have $M=S(b)/\sigma_0$ cells (like in the case with
local fusion) with $\eta_i$, $i=1,...,M$
fluctuated around $\beti$. Then (unlike the local fusion case)
we have to find
average $\eta_c=\frac{1}{M}\seta=\frac{N}{M}$ for given event,
and then to generate particles from one cluster
with average multiplicity equal to $\mu_c \sqrt{\eta_c}=$
$\mu_0 M \sqrt{\eta_c}$ $=\mu_0 M \sqrt{N/M}$ $=\mu_0\sqrt{MN}$.
The general formulae for this case was obtained in the section 5 of
\cite{vestn1}:
\beq
\avr{O} {\F}
=\frac{\sum_{\{\eta^{}_1,...,\eta^{}_M\}} \avr{O} {{\{\eta^{}_1,...,\eta^{}_M\}},\F} W(\Aeta)
p^{}_{\mu_c\sqrt{\frac{1}{M}\seta}} (\F)}
{\sum_{\{\eta^{}_1,...,\eta^{}_M\}} W(\Aeta)
p^{}_{\mu_c\sqrt{\frac{1}{M}\seta}} (\F)}
\label{avOcl}
\eeq
where $\mu_c = \mu_0 M $ with $M=S(b)/\sigma_0$.
The assumption $\beti>>1$ is essential, as only in this situation
we can consider that the transverse area
of the cluster $\Delta S$ is equal to all interaction area $S(b)$ ($b$-impact parameter).

The $\avr{O} {{\{\eta^{}_1,...,\eta^{}_M\}},\F}$ is the rates of the backward production from
configuration ${\{\eta^{}_1,...,\eta^{}_M\}}$. We have to use for $\avr{\B}{\F}$ correlations:
\beq
\avr{\B} {{\{\eta^{}_1,...,\eta^{}_M\}},\F} =\mu_c \sqrt{\eta_c}= \mu_0 M \sqrt{\frac{1}{M}\seta}
\label{Onncl}
\eeq
and for $\avr{\ppB}{\F}$ correlations:
\beq
\avr{\ppB} {{\{\eta^{}_1,...,\eta^{}_M\}},\F} = p^2 \sqrt{\eta_c}=p^2 \sqrt{\frac{1}{M}\seta}
\label{Optncl}
\eeq
We see that the difference with the case of local fusion
consists in replace
$\frac{1}{M}\reta\to\sqrt{\frac{1}{M}\seta}$.
As a consequence calculations
in the case of global string fusion are much more simple,
as we can reduce all sums $\sum_{\{\eta^{}_1,...,\eta^{}_M\}}$
to one sum $\sum_N$, as in the no fusion case
(see section 3 in \cite{vestn1}).
So in the global fusion case we can write simple formulas:
\beq
\avr{\B}{\F}
=\frac{\mu_0 \sqrt{M}
\sum_N \sqrt{N} W(N) p^{}_{\mu_0\sqrt{M} \sqrt{N}}(\F)}
        {\sum_N W(N) p^{}_{\mu_0\sqrt{M} \sqrt{N}}(\F)}
\label{BFcl}
\eeq
and
\beq
\avr{\ppB}{\F}
=\frac{p^2}{\sqrt{M}}
\frac{\sum_N \sqrt{N} W(N) p^{}_{\mu_0\sqrt{M} \sqrt{N}}(\F)}
     {\sum_N W(N) p^{}_{\mu_0\sqrt{M} \sqrt{N}}(\F)}
\label{pBFcl}
\eeq
where $W(N)$ is given by the formula
\beq
W(N) = \sum_{{\{\eta^{}_1,...,\eta^{}_M\}}} \delta_{N,\seta} \prod_i w (\eta_i)
\label{WN}
\eeq

We see that in the case of global fusion
(one cluster at large $\beti$ with area
$\Delta S$ being equal to all interaction area $S(b)$)
$n$--$n$ and $pt$--$n$ correlations are connected
\beq
\avr{\ppB}{\F}=\frac{p^2}{\mu_0 M} \avr{\B}{\F}
\hs{1} {\rm or} \hs{1}
\frac{\avr{\B}{\F}}{\av{\B}}
=\frac{\avr{\ppB}{\F}}{\av{\ppB}}
\label{pnB}
\eeq
Note that unlike the local fusion case
in this case we find this result without any assumptions
on the properties of $p(\F)$ and $w(\eta_i)$ distributions
and for arbitrary (even nonequal) $\beti$.

Clear that in this case
the results can depend only on mean number of strings $\bN$
and on combination $\mu_M$:
\begin{equation}\label{muM}
\bN=\sum_i \beti          \hs{2}        \mu_M\equiv\mu_0 \sqrt{M}
\end{equation}

Below we'll calculate numerically the correlation functions on
formulas (\ref{BFcl}) and (\ref{pBFcl}), but at first we would like
to find explicit formulas for global fusion case in Gauss approximation.
We'll see that results really depend only
on one combination of the variables
(\ref{muM}), namely on
\begin{equation}\label{scal}
\frac{\mu_M}{\sqrt{\bN}}=\frac{\mu_0}{4\sqrt{\bN/M}}
\end{equation}
and the $scaling$ takes place
as in the case with local fusion.


\section{Gauss approximation for {\it global} fusion at large
 string density.}

Acting as in the no fusion case in section 4 of \cite{vestn1}
(see also calculations in the Gauss approximation
in the section 3 of the present paper)
we find
\beq
\avr{\B}{\F} =\mu_M \sqrt{N^*}=\mu_0 \sqrt{M} \sqrt{N^*}
\label{BF2cl}
\eeq
or keeping in mind (\ref{pnB})
\beq
\frac{\avr{\B}{\F}}{\av{\B}}
=\frac{\avr{\ppB}{\F}}{\av{\ppB}}=\sqrt{\frac{N^*}{\bN}}
\label{BF3cl}
\eeq
The $N^*$ is the value of $N$ at which the function
\beq
\varphi(N,\F)=\frac{(N-\bN)^2}{2\bN(1-\lambda_\eta)}+
\frac{(\F-\mu_M \sqrt{N})^2}{2\mu_M \sqrt{N} (1-\lambda)}
\label{phicl}
\eeq
gains its minimum. In short notations
\beq
z\equiv\sqrt{\frac{N^*}{\bN}}, \hs{1}
f=\frac{\F}{\mu_M \sqrt{\bN}}=\frac{\F}{\av{\F}}, \hs{1}
a=\frac{\mu_M\kappa}{4\sqrt{\bN}}
=\frac{\mu_0\kappa}{4\sqrt{\bN/M}}
\label{notcl}
\eeq
we find the equation
\beq
z^3-z=a \left( \frac{f^2}{z^2}-1 \right)
\label{zcl}
\eeq
which defines the function $z=z(f)$ and then
using (\ref{BF3cl}) we can calculate correlation functions
\beq
\frac{\avr{\B}{\F}}{\av{\B}}
=\frac{\avr{\ppB}{\F}}{\av{\ppB}}=z\left(\frac{\F}{\av{\F}}\right)=z(f)
\label{BF4cl}
\eeq
and correlation coefficients for the global fusion case
\beq
\ol{\beta} =\ol{b}=z'(1)=\frac{a}{a+1}
=\frac{\mu_M\kappa}{\mu_M\kappa+4\sqrt{\bN}}
=\frac{\mu_0\kappa}{\mu_0\kappa+4\sqrt{\bN/M}}
\label{b4cl}
\eeq
We see again that in Gauss approximation
there is the same remarkable $scaling$.
The $pt$--$n$ and $n$--$n$ correlations depend only on
one combination of parameters: $a=\frac{\mu_0\kappa}{4\sqrt{\bN/M}}$.
Note that unlike the local fusion case
in this case we find this result for arbitrary (even not equal) $\beti$.

In the homogeneous situation all $\beti=\eta$ and we have
\beq
\bN=M\eta, \hs{1}
a=\frac{\mu_0\kappa}{4\sqrt{\eta}}
\label{not1cl}
\eeq
and
\beq
\ol{\beta} =\ol{b}=
\frac{\mu_0\kappa}{\mu_0\kappa+4\sqrt{\eta}}=\frac{a}{a+1}
\label{b5cl}
\eeq

We see that in the $homogeneous$ situation in Gauss approximation
the results with $local$ and $global$ fusion coincide
(as we have expected in section 5 of \cite{vestn1}).
We have the same equations (\ref{zA}) and (\ref{zcl})
with the same value of parameter $a$ (\ref{not1A}) and (\ref{not1cl}).
Note that in the no fusion case we have had {\it very different} equation
for $z(f)$ (see section 4 of \cite{vestn1}).

Unlike the local fusion case in the global fusion case
we can control the validity of Gauss approximation
making calculations on exact formulas (\ref{BFcl}) and (\ref{pBFcl})
at different values of $M$.

\begin{figure}[t]
\centerline{\epsfig{file=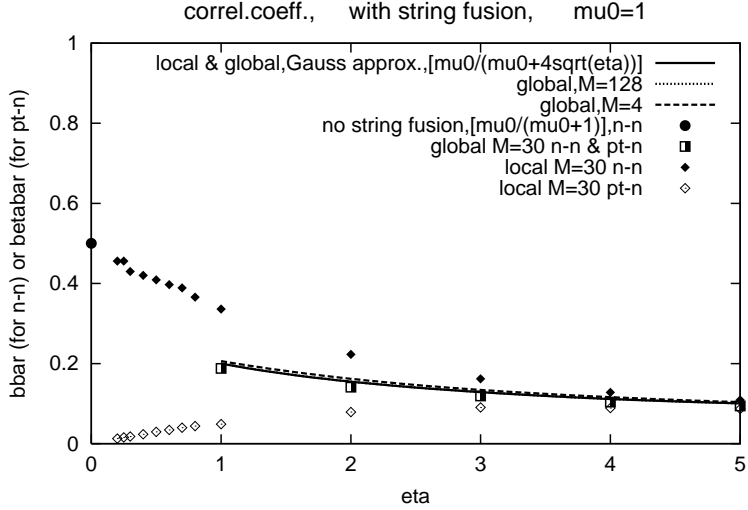,width=10cm,angle=-90}}
\caption[dummy]{\label{b1}
The $p_t$--$n$ and $n$--$n$ correlation coefficients at $\mu_0=1$.
Solid line - the Gauss approximation for local and global fusion
$\olb=\olbe=\mu_0/(\mu_0+4\sqrt{\eta})$.
Dashed and dotted lines - the results of exact calculations
on the formulas (\ref{BFcl}-\ref{pBFcl})
in the global fusion case at different values of $M$
(half-filled squares - the same by means of
direct numerical MC calculations on formulas (\ref{avOcl}-\ref{Optncl})).
The empty and filled points -
the results of direct numerical MC calculations
in the local fusion case based on formulas (\ref{avO}-\ref{Optn})
for the $pt$--$n$ and $n$--$n$ correlations correspondingly.
The filled circle - the $n$--$n$ correlation coefficient
$\olb=\mu_0/(\mu_0+1)$
in the case without string fusion \cite{vestn1}.
}
\end{figure}
\begin{figure}[t]
\centerline{\epsfig{file=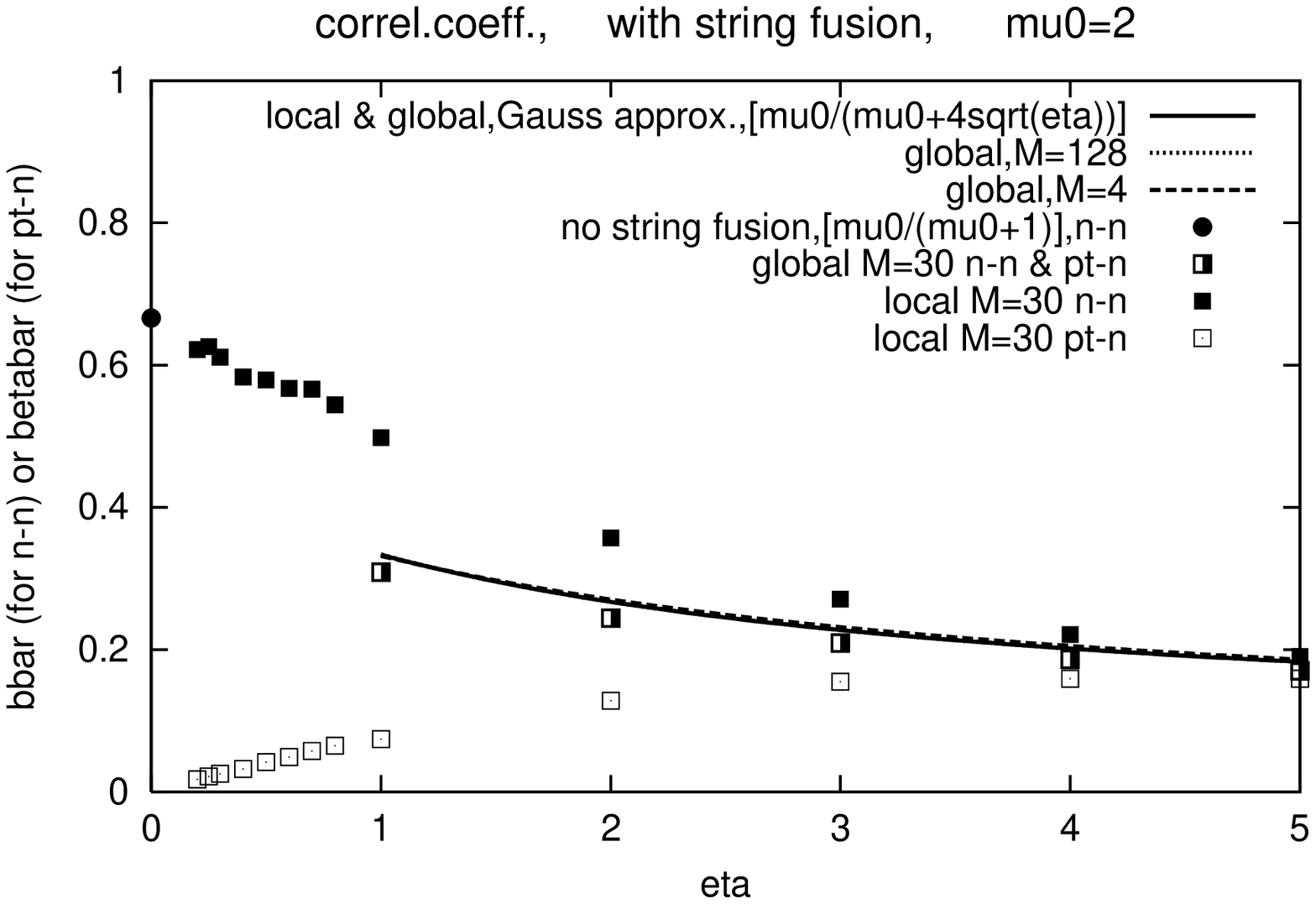,width=10cm,angle=-90}}
\caption[dummy]{\label{b2}
The same as in Fig.~\ref{b1}, but at $\mu_0=2$.
}
\end{figure}
\begin{figure}[t]
\centerline{\epsfig{file=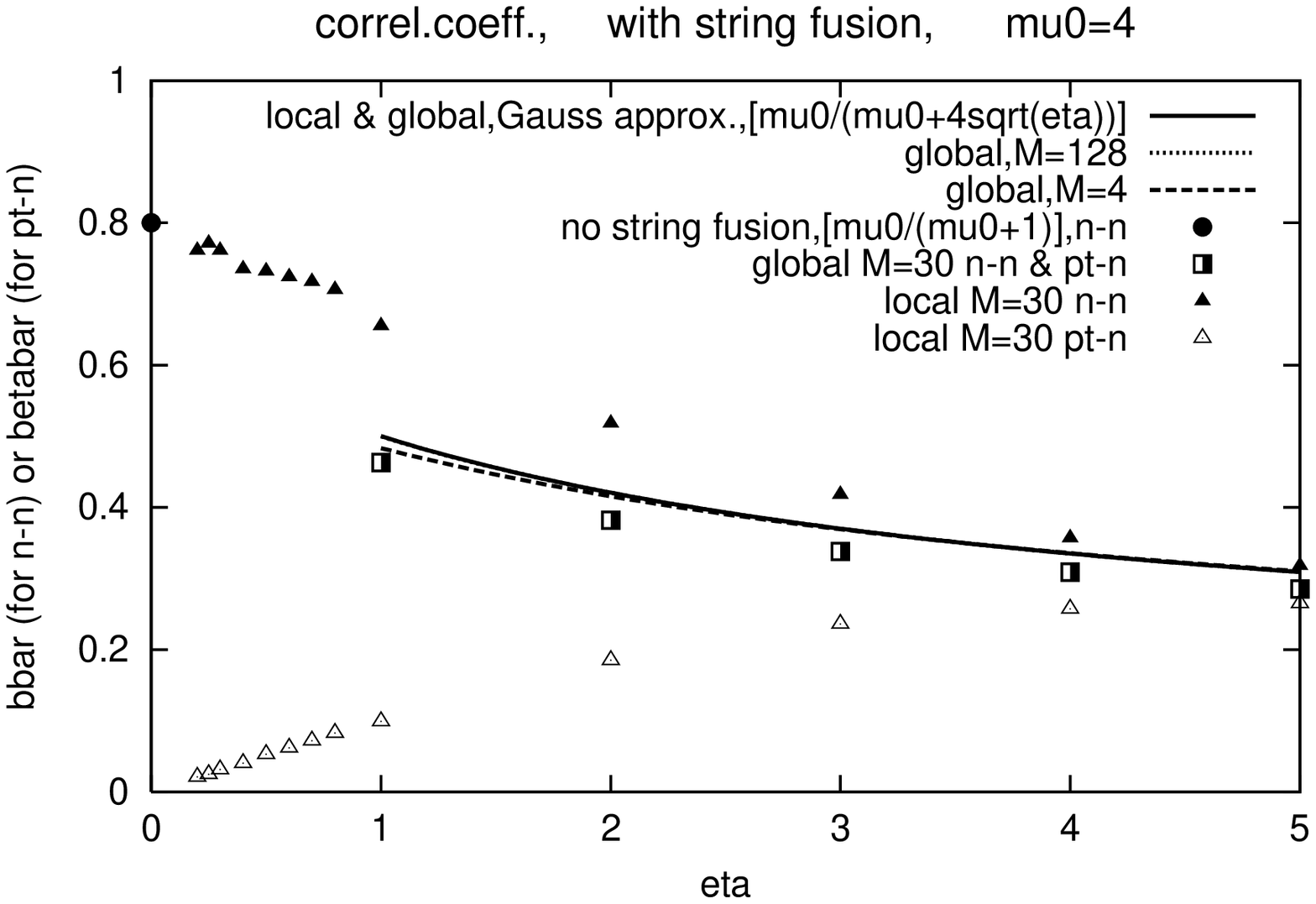,width=10cm,angle=-90}}
\caption[dummy]{\label{b4}
The same as in Fig.~\ref{b1}, but at $\mu_0=4$.
}
\end{figure}

Along with
the correlation function $z(f)$ (\ref{pnA}), (\ref{BF4cl}) in
Figs.~\ref{zf1},\ref{zf2}
and the correlation coefficient
$\ol{\beta}=\ol{b}$ (\ref{bb3A}),(\ref{b4cl})
in Figs.~\ref{b1}-\ref{b4}, calculated on our scaling formulas,
which in Gauss approximation are the same for local and global fusion
(the solid lines),
we present at the same pictures the results of exact calculations
in the global fusion case on formulas (\ref{BFcl}) and (\ref{pBFcl})
at different values of $M$ (the dotted and dashed lines).

We present also in Figs.~\ref{b1}-\ref{b4}
the results of our direct numerical MC calculations
of the $pt$--$n$ and $n$--$n$ correlation coefficients
in the global fusion case based on formulas (\ref{avOcl}-\ref{Optncl})
(half-filled squares).

We see that in this case the Gauss approximation works very well
and the $\mu_0/\sqrt{\eta}$-scaling is not an artifact of
this approximation.
More over along with the $\mu_0/\sqrt{\eta}$-scaling at large $\eta$
we have also $M$-independence for correlation
coefficients $\ol{\beta}$ and $\ol{b}$ (see Figs.~\ref{b1}-\ref{b4})
starting very early (from $M=4$).


\section{Conclusions.}

In conclusion let us compare the results obtained
in the present paper in the case with string fusion
with results obtained
in \cite{vestn1} in the case without string fusion.
\vskip 4mm

We have obtained in the paper \cite{vestn1}
in the case {\bf without} string fusion:
\begin{enumerate}
\item {for $n$--$n$ correlations:
\ $\ol{b} = \frac{a}{a+1}$ with $a=\mu_0\kappa$}
\item {for $pt$--$n$ correlations:
\ $\ol{\beta} =0$}
\end{enumerate}
\vskip 2mm

In the present paper in the case with the {\bf global} string fusion
and in the {\bf local} fusion case
for {\bf homogeneous} situation ($\beti=\eta$)
we find at large $\eta$:
\begin{enumerate}
\item {for $n$--$n$ correlations:
\ $\ol{b} = \frac{a}{a+1}$ with $a=\frac{\mu_0\kappa}{4\sqrt{\bN/M}}
                                         =\mu_0\kappa/(4\sqrt{\eta})$}
\item {for $pt$--$n$ correlations:
\ $\ol{\beta} =\ol{b} = \frac{a}{a+1}$ with the same $a$}
\end{enumerate}
We see that with fusion the $n$--$n$ correlations became weaker,
but now as compensation
we have the $pt$--$n$ correlations of the same strength.
We see also $\mu_0/\sqrt{\eta}$-scaling in this case.
\vskip 4mm

For {\bf nonhomogeneous} situation (different $\beti$)
in the case with {\bf local} string fusion
we have find at large $\beti$:
\begin{enumerate}
\item {for $n$--$n$ correlations:
\ $\ol{b} = \frac{a}{a+1}$ with $a=\mu_0\kappa/(4\br/M)$}
\item {for $pt$--$n$ correlations:
\ $\ol{\beta} =
\left( \frac{2\br^2}{\bN M}-1  \right) \ol{b}$}
and hence $\ol{\beta}\leq\ol{b}$
\end{enumerate}
$$
{\rm with} \hs{1} \bN=\sum_{i=1}^M \beti \hs{1}  {\rm and}  \hs{1} \br=\sum_{i=1}^M \sqrt{\beti}
$$
As we have demonstrated above
(see (\ref{bbeta2A}) and (\ref{bbeta3A})),
this leads to $\ol{\beta}$ smaller than $\ol{b}$: \ $\ol{\beta}\leq\ol{b}$.
It's possible situation (\ref{bbeta4A}), in which $ \ol{\beta} < 0 $.
\vskip 4mm

At {\bf small} string density as it's shown in Appendix A
the two types of limit at $\eta\to 0$ can be studied.
\begin{enumerate}
\item {If one keeps $M=const$, then we have $N \to 1$
(because the configurations with $N=0$, are not considered
as events) and  we have nor $p_t$--$n$ nor $n$--$n$ correlation.}
\item {If one keeps $\olN=const$, then $M\eta=const$ and $M\to\infty$,
hence the strings will be far separated in transverse plane and
we'll have the same results as in the case without
string fusion \cite{vestn1}.}
\end{enumerate}
\vskip 2mm

Note that the results obtained in our cellular approach
are in a good agreement with the results obtained
in the framework of the real string fusion model
taking into account detail geometry of strings overlapping
\cite{MAB}.


\subsection*{Acknowledgments.}

The authors would like to thank M.A.~Braun
and G.A.~Feofilov for numerous valuable
encouraging discussions.
The work has been partially supported by
the Russian Foundation for Fundamental Research
under Grant No. 01-02-17137-a.

\section*{Appendixes:}
\begin{appendix}
\section{Correlations at small string density.}
\label{ap:a}

In this appendix we calculate the correlation functions
and the correlation coefficients at small string density
in the case with local string fusion.
(In the case with global fusion this limit has no physical
sense (see discussion in the section 5 of \cite{vestn1})).

For clearness we consider
"homogeneous" case, when all $\beti$
is equal each other in the interaction area $\beti=\eta$.
Then in each cell $i$ ($i=1,...,M$) the $\eta_i$ fluctuate
around this mean value according to Poisson law
($\rho^{}_{a}(x)$ is the Poisson distribution with $\olx=a$):
\beq
w(\eta _i)=\rho _\eta ^{}(\eta _i)\equiv
e^{-\eta }\frac{(\eta )^{\eta _i}}{\eta _i!}
\label{ap:w}
\eeq

We'll assume also the Poissonian form of $p^{}_{\eta_i} (n_i)$:
\beq
p^{}_{\eta_i} (n_i) =  \rho^{}_{\mu_0\etai}(n_i)
 \equiv e^{-\mu_0\etai}\frac{(\mu_0\etai)^{n_i}}{n_i!}
\label{ap:Poisson}
\eeq
then we have the Poissonian distribution for
\beq
P^{}_{\{\eta^{}_1,...,\eta^{}_M\}} (\F) = \rho^{}_{\mu_0\reta}(\F)
\label{ap:Poisso}
\eeq
with $\avr{\F} {{\{\eta^{}_1,...,\eta^{}_M\}}} =
\mu_0 \reta = \mu_0 r = \avr{\F} {r}$.

Then we find from (\ref{avO}) for $n$--$n$ correlations:
\beq
\langle n_B^{}\rangle _{n_F^{}}^{}=
\frac{\mu _0\sum'_{\{\eta _1^{},...,\eta _M^{}\}}
r\left( \prod_iw(\eta _i)\right) \rho _{\mu _0r}^{}(n_F^{})}
{\sum'_{\{\eta _1^{},...,\eta _M^{}\}}\left( \prod_iw(\eta _i)\right)
\rho _{\mu _0r}^{}(n_F^{})}
\label{ap:nn}
\eeq
and for $pt$--$n$ correlations:
\beq
\langle p_{tB}^{2} \rangle _{n_F^{}}^{}=
\frac{p^2\sum'_{\{\eta _1^{},...,\eta _M^{}\}}
\frac{N}{r}\left( \prod_iw(\eta _i)\right) \rho _{\mu _0r}^{}(n_F^{})}
{\sum'_{\{\eta _1^{},...,\eta _M^{}\}}\left( \prod_iw(\eta _i)\right)
\rho _{\mu _0r}^{}(n_F^{})}
\label{ap:ptn}
\eeq
Recall that
$$
N=\sum_{i=1}^M \eta_i
                       \hs{1}{\rm and}\hs{1}
r=\sum_{i=1}^M \sqrt{\eta_i}
$$
The $N$ is the number of strings in the given event.

One has to omit in all $M$-fold sums
in (\ref{ap:nn}) and (\ref{ap:ptn}) one term, when all $\eta_i=0$,
which corresponds to the absence of inelastic interactions between
the nucleons of the colliding nuclei (we denote this fact
by $\sum'$).

The probability $P(n_F)$ to detect $n_F$ particles
in the forward rapidity window, which enters denominators
of (\ref{ap:nn}) and (\ref{ap:ptn}), is
\beq
P(n_F)=C{\sum_{\{\eta _1^{},...,\eta _M^{}\}}}^{\!\!\!\!\!\!\prime}
\;\; \left( \prod_i w(\eta _i)\right)
\rho _{\mu _0^{}r}^{}(n_F^{})
\label{ap:PnF}
\eeq
where from normalization condition we have
\beq
C=\frac{1}{1-w^M(0)}=
\frac{1}{1-e^{-M\eta}}
\label{ap:C}
\eeq

Clear that this factor $C$ is canceling in
the numerator and in the denominator
of (\ref{ap:nn}) and (\ref{ap:ptn}),
but if we calculate
the mean number of strings $\bN$,
we find
\beq
\olN=
C{\sum_{\{\eta _1^{},...,\eta _M^{}\}}}^{\!\!\!\!\!\!\prime}
\;\; \left( \prod_i w(\eta _i)\right) N =
C{\sum_{\{\eta _1^{},...,\eta _M^{}\}}}^{\!\!\!\!\!\!\prime}
\;\;
\left( \prod_i w(\eta _i)\right)
\left( \sum_i \eta _i \right) = \frac{M\eta}{1-e^{-M\eta}}
\label{ap:avN}
\eeq
and for the $\av{\F}$
at small $\eta\ll 1$
we have
\beq
\av{\F}=\sum_{n^{}_F} n^{}_F P(n^{}_F)=
C\mu^{}_{0F}{\sum_{\{\eta _1^{},...,\eta _M^{}\}}}^{\!\!\!\!\!\!\prime}
\;\; \left( \prod_i w(\eta _i)\right) r =
\mu^{}_{0F}\frac{M\eta}{1-e^{-M\eta}}=
\mu^{}_{0F}\olN
\label{ap:avF}
\eeq
Because for any $\omega>0$ we have $\sum_{\eta_i}^{}\eta_i^\omega
w(\eta_i)=\eta+O(\eta^2)$ at $\eta\to 0$, as the main contribution
comes from the term $\eta_i=1$.

There are two possibilities when $\eta\to 0$:
\begin{enumerate}
\item {$M=const$ and then $M\eta\to 0$ and $\olN\to 1$
(see (\ref{ap:avN}))}
\item {$\olN=const$ and then $M\eta=const$ ((see (\ref{ap:avN}))
and $M\to\infty$}
\end{enumerate}
We'll investigate both these possibilities.

The first one means that in the limit we have $N=\olN=1$
(because the configurations with $N=0$, are not considered
as events). Clear that in this situation we'll have
nor $p_t$--$n$ nor $n$--$n$ correlation, as we'll
have no fluctuations in the number of strings
(see discussion in the end of the section 4 of \cite{vestn1}).
Detail calculations are presented below.

In the second case with $\olN=const$ we'll have the
fluctuations in the number of strings $N$,
but in the limit $\eta\to 0$ we'll have $M\to\infty$,
the strings will be far separated
in transverse plane and the strings fusion will plays no role.
So in this case we'll have the same results as in
the no fusion case, considered in \cite{vestn1}:
large $n$--$n$ correlation
with a correlation coefficient equal to $b=\mu_0^{}/(\mu_0^{}+1)$
and no $p_t$--$n$ correlation.
(See calculations below.)

{\bf Detail calculations.}
Let us evaluate the $M$-fold sums
in (\ref{ap:nn}) and (\ref{ap:ptn}) at $\eta\to 0$
keeping all terms of order $(M\eta)^k$, $(M\eta)^k\eta$
with any $k$
and omitting all terms of order $(M\eta)^k\eta^2$ and higher.

The terms of order $(M\eta)^k$ originates from the
summands in (\ref{ap:nn}) and (\ref{ap:ptn})
with $\eta_{i_1}=...=\eta_{i_k}=1$ and other $\eta_{i}=0$.
The terms of order $(M\eta)^k\eta$ originates from the
summands in (\ref{ap:nn}) and (\ref{ap:ptn})
with $\eta_{i_1}=2$,
$\eta_{i_2}=...=\eta_{i_k}=1$ and other $\eta_{i}=0$.
Keeping this into mind we find for $P(n_F)$:
\beq
P(n_F)=C(G_0 + \frac{\eta}{2} G_1)
\label{ap:PnF1}
\eeq
and for $n$--$n$ correlations:
\beq
\langle n_B^{}\rangle _{n_F^{}}^{}=\mu^{}_{0B}
\frac{N_0 + \frac{\eta}{2} N_1}{G_0 + \frac{\eta}{2} G_1}
\label{ap:nn1}
\eeq
For $pt$--$n$ correlations we have:
\beq
\langle p_{tB}^{2} \rangle _{n_F^{}}^{}=p^2
\frac{P_0 + \frac{\eta}{2} P_1}{G_0 + \frac{\eta}{2} G_1}=
p^2 \left( 1 + \frac{\eta}{2} \frac{P_1-G_1}{G_0} \right)
\label{ap:ptn1}
\eeq
Here
\beq
G_0=P_0=\sum_{k=1}^{M} C_M^k \eta^k \rho^{}_{\mu_0 k}(n^{}_F)
\label{ap:coeff1}
\eeq
$$
N_0=\sum_{k=1}^{M} k C_M^k \eta^k \rho^{}_{\mu_0 k}(n^{}_F)
$$
$$
G_1=\sum_{k=1}^{M} k C_M^k \eta^k \rho^{}_{\mu_0 (k+\gamma)}(n^{}_F)
$$
$$
N_1=\sum_{k=1}^{M} (k+\gamma)k C_M^k \eta^k \rho^{}_{\mu_0 (k+\gamma)}(n^{}_F)
$$
$$
P_1=\sum_{k=1}^{M} \frac{k+1}{k+\gamma} k C_M^k \eta^k \rho^{}_{\mu_0 (k+\gamma)}(n^{}_F)
$$
where $\mu^{}_0\equiv\mu^{}_{0F}$ and $\gamma=\sqrt{2}-1$.

Note that at $M \gg 1$ and $M\eta=const$ we have for $P(n_F)$:
\beq
P(n_F)=C e^{-M\eta}\frac{\mu_0^{n^{}_F}}{n^{}_F !}
(\olG_0 + \frac{\eta}{2} e_{}^{-\mu^{}_0 \gamma} \olG_1)
\label{ap:PnF2}
\eeq
and for $n$--$n$ correlations:
\beq
\langle n_B^{}\rangle _{n_F^{}}^{}=\mu^{}_{0B}
\frac{\olN_0 + \frac{\eta}{2} e_{}^{-\mu^{}_0 \gamma} \olN_1}
{\olG_0 + \frac{\eta}{2} e_{}^{-\mu^{}_0 \gamma} \olG_1}
\label{ap:nn2}
\eeq
For $pt$--$n$ correlations we have:
\beq
\langle p_{tB}^{2} \rangle _{n_F^{}}^{}=p^2
\frac{\olP_0 + \frac{\eta}{2} e_{}^{-\mu^{}_0 \gamma} \olP_1}
{\olG_0 + \frac{\eta}{2} e_{}^{-\mu^{}_0 \gamma}\olG_1}=
p^2 \left( 1 + \frac{\eta}{2} e_{}^{-\mu^{}_0 \gamma}
\frac{\olP_1-\olG_1}{\olG_0} \right)
\label{ap:ptn2}
\eeq
Here
\beq
\olG_0=\olP_0=\sum_{k=1}^{\infty} k_{}^{n^{}_F} \frac{d^k}{k!}
\label{ap:coeff2}
\eeq
$$
\olN_0=\sum_{k=1}^{\infty} k_{}^{n^{}_F+1} \frac{d^k}{k!}
$$
$$
\olG_1=\sum_{k=1}^{\infty} (k+\gamma)_{}^{n^{}_F} \frac{d^k}{(k-1)!}
$$
$$
\olN_1=\sum_{k=1}^{\infty} (k+\gamma)_{}^{n^{}_F+1} \frac{d^k}{(k-1)!}
$$
$$
\olP_1=\sum_{k=1}^{\infty} (k+\gamma)_{}^{n^{}_F-1}(k+1) \frac{d^k}{(k-1)!}
$$
where $d=M\eta e^{-\mu^{}_0}$.

For the control of calculations
we have used also the following explicit formulas at
$n^{}_F=0,1,2$:
\beq
\olG_0(0)=\olP_0(0)=e^d-1
\label{ap:coeff3}
\eeq
$$
\olN_0(0)=\olG_1(0)=\olG_0(1)=\olP_0(1)=de^d
$$
$$
\olN_1(0)=\olG_1(1)=(d+\sqrt{2})de^d
$$
$$
\olP_1(0)=\sum_{k=1}^{\infty} \frac{k+1}{k+\gamma}\; \frac{d^k}{(k-1)!}
$$
$$
\olN_0(1)=\olG_0(2)=\olP_0(2)=(d+1)de^d
$$
$$
\olN_1(1)=\olG_1(2)=(d^2+d(1+2\sqrt{2})+2)de^d
$$
$$
\olP_1(1)=(d+2)de^d
$$
$$
\olN_0(2)=(d^2+3d+1)de^d
$$
$$
\olN_1(2)=(d^3+d^23(1+\sqrt{2})+d(7+3\sqrt{2})+2\sqrt{2})de^d
$$
$$
\olP_1(2)=(d^2+d(3+\sqrt{2})+2\sqrt{2})de^d
$$

\begin{figure}[t]
\centerline{\epsfig{file=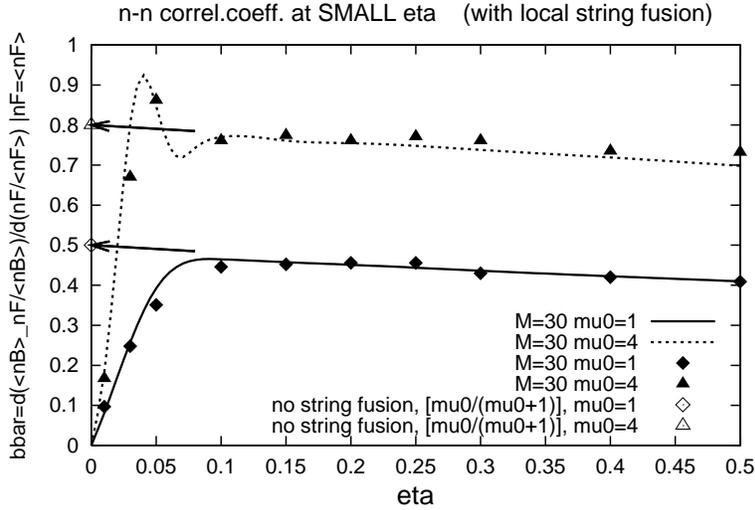,width=10cm,angle=-90}}
\caption[dummy]{\label{sm1}
The $n$--$n$ correlation coefficient
at small values of $\eta$ for $\mu_0=1$ and $\mu_0=4$.
The lines - results of the calculations using the asymptotic formulas
(\ref{ap:nn1}) and (\ref{ap:coeff1}) at $M=const$ ($\olN\to 1$).
The points - results of direct MC numerical calculations using
formula (\ref{ap:nn}).
The arrows show the value of the $n$--$n$ correlation coefficient
$\olb=\mu_0/(\mu_0+1)$
in the case without string fusion \cite{vestn1},
which corresponds to the limit $\olN=const$ ($M\to\infty$).
}
\end{figure}
\begin{figure}[t]
\centerline{\epsfig{file=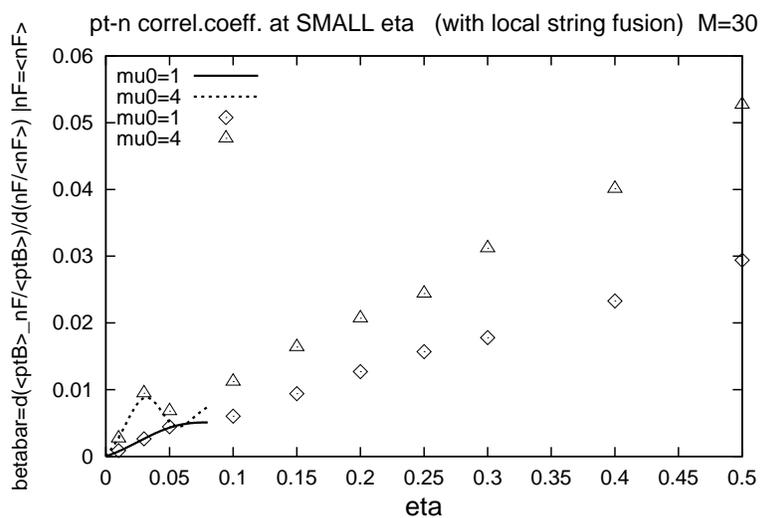,width=10cm,angle=-90}}
\caption[dummy]{\label{sm2}
The $p_t$--$n$ correlation coefficients
at small values of $\eta$ for $\mu_0=1$ and $\mu_0=4$.
The lines - results of the calculations using the asymptotic formulas
(\ref{ap:ptn1}) and (\ref{ap:coeff1}) at $M=const$ ($\olN\to 1$).
The points - results of direct MC numerical calculations using
formula (\ref{ap:ptn}).
}
\end{figure}

The results of the calculations using the asymptotic formulas
(\ref{ap:nn1}--\ref{ap:coeff1})
in the first case
($M=const$, $\eta\to 0$, $M\eta\to 0$ and $\olN\to 1$)
are shown in the Figs.~\ref{sm1}-\ref{sm2} (lines) together
with results of direct MC numerical calculations using
formulas (\ref{ap:nn}) and (\ref{ap:ptn})(points).
As we have expected in this case both
$n$--$n$ and $p_t$--$n$ correlation coefficients
go to zero when $\olN\sim 1$, i.e. at $\eta<1/M$.
Remember that we have $M$-independence
for the correlation coefficients at large $\eta$.
Now we see
that in this limit it disappears at $\eta \leq 1/M$.
This is also the reason for nonlinear dependence
of the correlation coefficients on $\eta$  in this region
which one can see in Figs.~\ref{sm1}-\ref{sm2} for $\mu_0=4$.

We see also that
using asymptotic formulas (\ref{ap:nn1}--\ref{ap:coeff1})
we can calculate
$n$--$n$ correlation coefficients
in wider region of small $\eta$, than
$p_t$--$n$ correlation coefficients,
because we have the contributions of order
$(M\eta)^k$ and $(M\eta)^k\eta$ for
$n$--$n$ correlations and only first non-trivial
contribution of order $(M\eta)^k\eta$ for
$p_t$--$n$ correlations.
Note the very good agreement between results of
the calculations on the asymptotic formulas
(\ref{ap:nn1}--\ref{ap:coeff1})
(lines in the Figs.~\ref{sm1}-\ref{sm2})
and
the results of direct MC numerical calculations on
formulas (\ref{ap:nn}) and (\ref{ap:ptn})
(points in the Figs.~\ref{sm1}-\ref{sm2}).


In the second case when $\eta\to 0$
we keep $\olN=const$ and then due to (\ref{ap:avN}) $M\eta=const$
so $M\to\infty$. We can use formulas
(\ref{ap:nn2}--\ref{ap:coeff2}) with $d=const$,
then in the limit $\eta\to 0$ we find for $n$--$n$ correlations:
\beq
\langle n_B^{}\rangle _{n_F^{}}^{}=
\mu^{}_{0B} \frac{\olN_0}{\olG_0}=
\mu^{}_{0B}
\frac{\sum_{k=1}^{\infty} k \rho_{M\eta}^{}(k) \rho_{\mu _0 k}^{}(n_F^{})}
{\sum_{k=1}^{\infty} \rho_{M\eta}^{}(k) \rho_{\mu _0 k}^{}(n_F^{})}
\label{ap:nn3}
\eeq
Here we multiply both the numerator and the denominator by
$e^{-M\eta }\mu _0^{n_F^{}}/n_F^{}!$.

Recall that
$\rho^{}_{a}(x)$ is the Poisson distribution with $\olx=a$,
then we see that formula (\ref{ap:nn3}) coincides with
the formula for $n$--$n$ correlations obtained in the paper
\cite{vestn1} in the case {\it without} string fusion
(see formula (23) in \cite{vestn1}).
So we'll have the same result for the $n$--$n$
correlation coefficient $b=\mu_0^{}/(\mu_0^{}+1)$ as in the
no fusion case (see Figs.~\ref{sm1}-\ref{sm2}).

For $p_t$--$n$ correlation coefficients
in this limit we find from
(\ref{ap:ptn2}) and (\ref{ap:coeff2}) with $d=const$:
\beq
\langle p_{tB}^{2}\rangle _{n_F^{}}^{}=
p^2 ( 1 + O(\eta))
\label{ap:ptn3}
\eeq
So we have no $p_t$--$n$ correlation as in the case without
string fusion \cite{vestn1}.

\section{
On the difference between $\av{\B}$ and $\avr{\B} {\F=\av{\F}}$.}
\label{ap:b}

It's possible instead of (\ref{bbA}) and (\ref{bbetaA})
to use the following definitions for the correlation coefficients:
\beq
\ol{b} \equiv\frac{d\avr{\B}{\F}/\avr{\B} {\av{\F}}}{d\F/\av{\F}} \at {\F=\av{\F}}
\label{A1}
\eeq
and
\beq
\ol{\beta} \equiv\frac{d\avr{\ppB}{\F}/\avr{\ppB} {\av{\F}}}{d\F/\av{\F}} \at {\F=\av{\F}}
\label{A2}
\eeq
where
\beq
\avr{\B}{\av\F} = \avr{\B}{\F=\av\F}
\label{A3}
\eeq
and
\beq
\avr{\ppB}{\av\F} = \avr{\ppB}{\F=\av\F}
\label{A4}
\eeq
Clear that
\beq
\av\B = \sum_\F \olP(\F) \avr{\B}{\F} \approx \avr{\B}{\F=\av\F} \sum_\F P(\F)
= \avr{\B}{\F=\av\F} \equiv \avr{\B}{\av\F}
\label{A5}
\eeq
and the same for $\avr{\ppB}{\av\F} \equiv  \avr{\ppB}{\F=\av\F} \approx \av{\ppB} $.

In our Gauss approximation these two types of quantities
coincide with each other.
\end{appendix}



\end{document}